\documentclass[12pt]{article}

\newcommand{\beg}{\begin{equation}}
\newcommand{\ene}{\end{equation}}

\begin{document}
\title{
\textsc{\bf Local Realism, Contextualism  and}\\
\textsc{\bf Loopholes in Bell`s Experiments}}

\author{
Andrei Khrennikov and Igor Volovich
\\ $~~~$\\
\textsf{International Center in Mathematical Modelling in }\\
\textsf{Physics, Engineering and Cognitive Sciences}\\
\textsf{University of Vaxjo, S-35195, Sweden}\\
\emph{e-mail: Andrei.Khrennikov@msi.vxu.se}\\
 $~~~$\\
\textsf{Steklov Mathematical Institute}\\
\textsf{Russian Academy of Sciences}\\
\textsf{Gubkin St. 8, 117966, GSP-1, Moscow, Russia}\\
\emph{e-mail: volovich@mi.ras.ru} }
\date {~}
\maketitle
\begin{abstract}
 It is currently  widely accepted, as a result of
 Bell's theorem and related experiments, that quantum mechanics is
 inconsistent with  local realism and there is the so called
 quantum non-locality. We show that such a claim  can be justified
 only in a simplified approach to quantum mechanics when one
 neglects the fundamental fact that there exist space and time.
 Mathematical definitions of local realism in the sense of Bell and
 in the sense of Einstein are given. We demonstrate that if we
 include into the quantum mechanical formalism the space-time
 structure in the standard way then quantum mechanics might be
 consistent with Einstein's local realism. It shows that loopholes
 are unavoidable in experiments aimed to establish a violation of
 Bell`s inequalities. We show how the space-time structure can be
 considered from the contextual point of view. A mathematical
framework for the contextual approach is outlined.
\end{abstract}
\newpage

\tableofcontents

%%%%%%%%%%%%%%%%%%%%%%%%%%%%%%%%%%%%%%%%%%%%%%%%%%%%

\section{Introduction}
\subsection{ Quantum non-locality without space and time?}

 Einstein, Podolsky and Rosen (EPR) presented an argument to show
 that there are situations in which the scheme of quantum theory
 seems to be incomplete \cite{EPR}. They proposed a
gedanken experiment involving a system of two particles spatially
separated but correlated in position and momentum and argued that
two non-commuting variables (position and momentum of a particle)
can have simultaneous physical reality. They concluded that the
description of physical reality given by  quantum mechanics,
which, due to the uncertainty principle, does not permit such a
simultaneous reality, is incomplete.

Though the EPR work dealt with continuous position and momentum
variables most of the further activity have concentrated almost
exclusively on systems of discrete spin variables following to the
Bohm \cite{Boh} and Bell \cite{Bel1} works.

Entangled states, i.e. the states of two particles with the wave
function which is not a product of the wave functions of single
particles,  have been studied in many theoretical and experimental
works starting from works of Einstein, Podolsky and Rosen and
Schrodinger.

Bell's theorem~\cite{Bel1} states that there are quantum spin
correlation functions that can not be represented as classical
correlation functions of separated  random variables. It has been
interpreted as incompatibility of the requirement of locality with
the statistical predictions of quantum mechanics~\cite{Bel1}. For a
recent discussion of Bell's theorem see, for example ~\cite{CS} -
~\cite{Vol2} and references therein.
 It is now widely accepted, as a result of
Bell's theorem and related experiments, that " local realism" must
be rejected and there exists the  so called quantum non-locality.

However it was shown in \cite{Vol1, Vol2,Vol3} that in the
derivation of such a   conclusion the fundamental fact that
space-time exists was neglected. Moreover, if we take into account
the spatial dependence of the wave function then the standard
formalism of quantum mechanics might be  consistent with local
realism.

\subsection{ Contextual approach}

From the other side a general contextual approach to the
probabilistic scheme of  quantum theory was proposed in
\cite{Khr3}. It is based on the transformation rules induced by
context transitions.  Context is a complex of physical conditions
used for the preparation of quantum or classical states. The idea
of the contextual dependence of probabilistic results of
observations is a very general one. It can be used and developed
in various directions. In particular it was suggested in
\cite{Vol3} to treat boundary conditions for quantum mechanical
differential equations as an appropriate context.

Context describes a measure of idealization which we use to
construct a mathematical model for a physical process. For example
in some approximation one can deal with models of quantum
phenomena when the spatial characteristics are neglected as it was
done by Bell in his consideration of the EPR paradox. However if
we want to speak about fundamental properties of quantum theory
then the principal role of the space-time picture should  not be
overlooked. In axiomatic approach to quantum theory after von
Neumann \cite{Neu} one often postulates only the formalism of
Hilbert space, its statistical interpretation and the abstract
Schrodinger evolution equation but without indication to the
spatial properties of quantum system. The necessity of including
into the list of basic axioms of quantum mechanics the property of
covariance of the physical system under the spatial translation
and rotation and moreover under  the Galilei or Poincare group was
stressed in \cite{Vol4}.

 In this paper we combine the spatial
approach to problems of quantum non-locality from \cite{Vol2,Vol4}
with the contextual approach of \cite{Khr3} to investigate
problems of quantum non-locality and local realism. We  consider
the space-time as a context for a quantum model. We will present a
mathematical formalism for the contextual approach. We will give
also two different definitions of the notions of local realism
which we call Bell's and Einstein's local realism. We demonstrate
that  if we include into the quantum mechanical formalism the
space-time structure in the standard way then quantum mechanics
actually  is consistent with local realism. Since detectors of
particles are obviously located somewhere in space  it shows that
loopholes are unavoidable in experiments aimed to establish a
violation of Bell`s inequalities. Our main tool will be analysis
of correlation functions in quantum and in classical theory.

\subsection{ Bell's local realism}

A mathematical formulation of Bell's local realism may be given by
relation (in more details it is discussed in the next section)
\begin{equation}
 \left<\psi |A(a)B(b)|\psi \right>=E\xi (a)\eta
(b) \label{eq:eqni1}
\end{equation}
Here $A(a)$ and $B(b)$ are self-adjoint operators which commute on
a natural domain and $a$ and $b$ are certain indices. Here $E$ is
a mathematical expectation and $\xi (a)$ and $\eta (b)$ are two
stochastic processes and  $\psi$ is a vector from a Hilbert space.
Then we say that  the triplet
$$
\{A(a),B(b),\psi\}
$$
satisfies the {\it Bell`s local realism} (BLR) condition.

Bell proved that a two spin quantum correlation function which is
equal to just $-a\cdot b $, where $a$ and $b$ are two
3-dimensional vectors, can not be represented in the form
(\ref{eq:eqni1} ),
\begin{equation}
\left<\psi_{spin}|\sigma\cdot a \otimes\sigma\cdot
b|\psi_{spin}\right>\neq E\xi (a)\eta (b) \label{eq:eqni1a}
\end{equation}
if one has a bound $|\xi (a)|\leq 1$,~  $ | \eta (b)|\leq 1.$ Here
$a=(a_1,a_2,a_3)$ and $b=(b_1,b_2,b_3)$ are two unit vectors in
three-dimensional space ${\bf R}^3$ and
$\sigma=(\sigma_1,\sigma_2,\sigma_3)$ are the Pauli matrices,

Therefore the correlation function of two spins does not satisfy
to the BLR condition (\ref{eq:eqni1}). In this sense sometimes one
speaks about quantum non-locality.

\subsection{ Space and time in axioms of quantum mechanics}

Note however that in the previous discussion the space-time
parameters were not explicitly involved though one speaks about
non-locality. Actually the "local realism" in the Bell sense as it
was formulated above in Eq. (\ref{eq:eqni1}) is a  notion which
has nothing to do with notion of locality in the ordinary 3
dimensional space. Therefore we  define also another notion which
we will call the condition of local realism in the sense of
Einstein.

To explain the notion let us first remind that the usual axiomatic
approach to quantum theory involves only the Hilbert space,
observable, the density operator $\rho$ and the von Neumann
formula for the probability $P(B)$ of the outcome $B$:
$P(B)=Tr\rho E_B$ where $\{E_B\}$ is POVM associated with a
measured space $(\Omega,{\cal F}),$ here $B$ belongs to the
$\sigma$-algebra ${\cal F}$ . It was stressed in \cite{Vol4} that
in a more realistic axiomatic approach to quantum mechanics one
has to includes an axiom on the existing of space and time. It can
be formulated as follows
$$
U(d)E_BU(d)^*=E_{\alpha_d(B)}
$$
Here $U(d)$ is  the unitary representation of the group of
translations in time and in the three-dimensional space and
$\alpha_d:{\cal F}\to {\cal F}$ is the group of automorphisms..

\subsection{ Einstein's local realism}

 Let in  a Hilbert  space $\cal H$ be given a family of self-adjoint
operators $\{A(a,\cal O)\}$ and $\{B(b,{\cal O})\}$ parameterized
by the regions $\cal O$ in the Minkowsky space-time. Suppose that
one has a representation
\begin{equation}
\left<\psi |A(a,{\cal O}_1)B(b,{\cal O}_2)|\psi \right> =E\xi
(a,{\cal O}_1)\eta (b,{\cal O}_2) \label{eq:eqni1C}
\end{equation}
for  $a, b, {\cal O}_1, {\cal O}_2$ for which the operators
commute. Then we say that the quadruplet
$$
\{A(a,{\cal O}_1),B(b,{\cal O}_2),U(d),\psi\}
$$
 satisfies the {\it Einstein local realism} (ELR) condition.

\subsection{ Local realist representation for quantum spin
correlations}

Quantum correlation describing the localized measurements of spins
in the regions ${\cal O}_1$ and ${\cal O}_2$ includes the
projection operators $P_{{\cal O}_1}$ and $P_{{\cal O}_2}$ . In
contrast to Bell`s theorem (\ref{eq:eqni1a} ) now there exists a
local realist representation \cite{Vol2}
\begin{equation}
 \left<\psi| \sigma\cdot a P_{{\cal O}_1}\otimes
\sigma\cdot b  P_{{\cal O}_2} |\psi\right> =E\xi ({\cal O}_{1},a)
\eta({\cal O}_2,b)
 \label{eq:eqni1d}
\end{equation}
if the distance between the regions ${\cal O}_1$  and ${\cal O}_2$
is large enough. Here  all classical random variables are bounded
by 1.

Since detectors of particles are obviously located somewhere in
space  it shows that loopholes are unavoidable in experiments
aimed to establish a violation of Bell`s inequalities. Though
there were some reports on experimental derivation of violation of
Bell's inequalities, in fact such violations always were based on
additional assumptions besides local realism. No genuine Bell's
inequalities have been violated since always some loopholes were
in the experiments, for a review see for example \cite{CS,San}.
  There were many discussions of proposals for experiments which
could avoid the loopholes however up to now  a convincing proposal
still did not advanced .

One can compare the situation with attempts to measure the
position and momentum of a particle in a single experiment. Also
one could speak about some technical difficulties (similar to the
efficiency of detectors loophole) and hope that some could come
with a proposal to make an experiment without loopholes. However
we know from the uncertainty relation for the measurement of
momentum and position that it is not possible. Similarly the
formula (\ref{eq:eqni1d}) shows that a loophole free experiment in
which a violation of Bell's inequalities will be observed is
impossible if the distance between detectors is large enough.
Therefore loopholes in Bell's experiments are irreducible.

\subsection{ EPR versus Bohm and Bell}

The original EPR system involving continuous variables has been
considered by Bell in \cite{Bel2}. He has mentioned that if one
admits "measurement" of arbitrary "observable" at arbitrary state
than it is easy to mimic his work on spin variables (just take a
two-dimensional subspace and define an analogue of spin
operators). The problem which he was discussing in \cite{Bel2} is
narrower problem, restricted to measurement of positions only, on
two non-interacting spin-less particles in free space. Bell used
the Wigner distribution approach to quantum mechanics.The original
EPR state has a nonnegative Wigner distribution. Bell argues that
it gives a local, classical model of hidden variables and
therefore the EPR state should not violate local realism. He then
considers a state with non-positive Wigner distribution and
demonstrates that this state violates local realism.

Bell`s proof of violation of local realism in phase space has been
criticized in \cite{Joh} because of the use of an unnormalizable
Wigner distribution. Then in \cite{BW} it was demonstrated that
the Wigner function of the EPR state, though positive definite,
provides an evidence of the nonlocal character of this state if
one measures a parity operator.

In \cite{KhV2} we have applied to the original EPR problem the
method which was used by Bell in his well known paper \cite{Bel1}.
He has shown that the correlation function of two spins  cannot be
represented by classical correlations of separated bounded random
variables. This Bell`s theorem  has been interpreted as
incompatibility of local realism with quantum mechanics.  It was
shown in \cite{KhV2} that, in contrast to Bell`s theorem for spin
correlation functions, the correlation function of positions (or
momenta) of two particles always admits a representation in the
form of classical correlation of separated random variables.The
following representation was proved
\begin{equation}
\label{eq:2} \left<\psi|q_1 (\alpha_1) q_2(\alpha_2)|\psi\right>=E
\xi_1 (\alpha_1) \xi_2(\alpha_2) \end{equation} The explanation of
the notations see below. Therefore we obtain a local realistic (in
the sense of Bell and in the sense of Einstein as well)
representation for the correlation function in the original EPR
model.
  This result looks rather surprising
since one thinks that the Bohm-Bell reformulation of the
EPR paradox is equivalent to the original one.

\section{Correlation functions and local realism}

A mathematical formulation of Bell's local realism may be given as
follows. Let, in a Hilbert space $\cal H$, be given two families
of self-adjoint operators $\{A(a)\}$ and $\{B(b)\}$ which commute
$[A(a),B(b)]=0$ on a natural domain. Here $a$ and $b$ are elements
of two arbitrary sets of indices. Suppose that one has a
representation
\begin{equation}
\left<\psi |A(a)B(b)|\psi \right>=E\xi (a)\eta (b)
\label{eq:eqnii1}
\end{equation}
for any $a, b$ where $E$ is a mathematical expectation and $\xi
(a)$ and $\eta (b)$ are two stochastic processes such that the
range of $\xi (a)$ is the spectrum of $A(a)$ and the range of
$\eta (b)$ is the spectrum of $B(b)$. Here $\psi$ is a vector from
$\cal H$. Then we say that {\it the triplet
$$\{\{A(a)\},\{B(b)\},\psi\}$$ satisfies the BLR  (Bell`s local
realism) condition}.

Bell proved that a two spin quantum correlation function which is
equal to just $-a\cdot b $, where $a$ and $b$ are two
3-dimensional vectors, can not be represented in the form
(\ref{eq:eqn1} ) if one has a bound $|\xi (a)|\leq 1$,~  $ | \eta
(b)|\leq 1.$ Therefore the correlation function of two spins does
not satisfy to the BLR condition (\ref{eq:eqnii1}). In this sense
sometimes one speaks about quantum non-locality.

Note however that in the previous discussion the space-time
parameters were not explicitly involved though one speaks about
non-locality. Actually the "local realism" in the Bell sense as it
was formulated above in Eq. (\ref{eq:eqn1}) is a very general
notion which has nothing to do with notion of locality in the
ordinary three-dimensional space. We will define now another
notion which we will call the condition of local realism in the
sense of Einstein. First let us recall that in quantum field
theory the condition of locality (local commutativity) reads:
\begin{equation}
[F(x),G(y)]=0 \label{eq:eqn1A}
\end{equation}
if the space-time points $x$ and $y$ are space-like separated.
Here $F(x)$ and $G(y)$ are two Bose field operators (for Fermi
fields we have anti-commutator).

Let in the Hilbert space $\cal H$ there is a unitary
representation $U$ of the inhomogeneous Lorentz group and let be
given a family of self-adjoint operators $\{A(a,\cal O)\}$
parameterized by the regions $\cal O$ in Minkowsky space-time
where $a$ is an arbitrary index. Let us suppose that the unitary
operator translations act as
\begin{equation}
U(d)A(a,{\cal O})U(d)^*=A(a,{\cal O} (d)) \label{eq:eqn1B}
\end{equation}
where $d$ is a four dimensional vector and ${\cal O} (d))$ is a
shift of ${\cal O} $ at $d$. Let be given also a family of
operators $\{B(b,{\cal O})\}$ with similar properties. Suppose
that one has a representation
\begin{equation}
\left<\psi |A(a,{\cal O}_1)B(b,{\cal O}_2)|\psi \right> =E\xi
(a,{\cal O}_1)\eta (b,{\cal O}_2) \label{eq:eqn1C}
\end{equation}
for  $a, b, {\cal O}_1, {\cal O}_2$ for which the operators
commute
$$[A(a,{\cal O}_1),B(b,{\cal O}_2)]=0$$
The correlation function (\ref{eq:eqn1C}) describes the results of
a simultaneous measurement. Moreover we suppose that the range of
$\xi (a,{\cal O}_1)$ is the spectrum of $A(a,{\cal O}_1)$ and the
range of $\eta (b,{\cal O}_2)$ is the spectrum of $B(b,{\cal
O}_2)$. Then we say that {\it the quadruplet
$$\{\{A(a,{\cal O}_1)\},\{B(b,{\cal O}_2)\},U,\psi\}$$ satisfies
the ELR (Einstein local realism) condition}.

For Fermi fields which anti-commute we assume the same relation
(\ref{eq:eqn1C}) but the random fields $\xi$ and $\eta$ should be
now anti-commutative random fields (superanalysis and probability
with anticommutative variables are considered in \cite{VV, Khr4}).

One can write an analogue of the presented notions in the case
when the region $\cal O$ shrinks to a point (in such a case we
have an operator $A(a,x)$) and also for $n$-point correlation
functions
\begin{equation}
\left<\psi |A_1(a_1,x_1)...A_n(a_n,x_n)|\psi \right>
=E\xi_1(a_1,x_1)...\xi_n(a_n,x_n) \label{eq:eqn1D}
\end{equation}
A non-commutative spectral theory related with such representations
is considered in \cite{Vol3}.

\section{Contextual classical and quantum probability}

The contextual probabilistic approach is nothing than
probabilistic formalization of Bohr's idea that the whole
experimental arrangement must be taken into account. The basic
postulate of the contextual probabilistic approach to general
statistical measurements is that {\bf probability distributions
for physical variables depend on complexes of experimental
physical conditions.} Such complexes are called (experimental)
contexts. Mathematically contextualism means the impossibility to
operate with an abstract (e.g. Kolmogorov) probability ${\bf P}$
indending of a context. Thus, in the opposite to traditions in
probability theory, we could not work with e.g. a single
Kolmogorov probability space $(\Omega, {\cal F}, {\bf P})$ that
was fixed once and for ever. If we choose the measure theoretical
approach to probability (Kolmogorov, 1933), then in the
contextualist probabilistic framework we should to work with
families of Kolmogorov probability spaces. Here mathematically
every context is represented by its own probability space, compare
to the camelion approach of L. Accardi \cite{Accardi} and the theory of
probability manifolds of S. Gudder \cite{Gudder} .

 Moreover, Kolmogorov's
measure-theoretical probability theory \cite{Kolmogorov} is not so natural as the
mathematical base for the contextual probabilistic approach to
statistical measurements. The main contribution of A. N.
Kolmogorov into axiomatisation of probability theory was
consideration of abstract probability measures. The great
advantage of the Kolmogorov probability theory was the possibility
to perform general probabilistic derivations, i.e., derivations
for abstract probabilities without to take into account contextual
dependence of probabilities.

But in the contextual probabilistic framework it would be more
natural to start not with an abstract probability, but directly
with a context and then consider a sequence of experimental trails
in this context. As the result, we get a sequence of physical
characteristics of systems under consideration. Then we can define
the probability distribution (if it exists at all) of those
characteristics by using the {\it principle of statistical
stabilization} of relative frequencies. Mathematical
formalization of this approach ({\it frequency probability
theory})
 was proposed by R. von Mises \cite{Mises} on the basis of theory of
{\it collectives} (random sequences). Thus if we use the frequency
probability theory, then we can identify a context with a
collective. Our fundamental thesis  {\it ``first context -- then
probability distribution"} is closely related to von Mises'
fundamental thesis: {\it ``first collective
 -- then probability distribution".}

The authors of the paper are well aware that the original von
Mises definition of collective was not mathematically rigorous,
see e.g.  on the details. This problem induced extended
investigations on the notion of randomness, see e.g. \cite{Vit}, \cite{KHRA}. In
particular, those investigations induced the theory of recursive
functions and Kolmogorov's algorithmic complexity. We recall that,
in particular, if we restrict the class of von Mises place
selections to recursive functions, then we get  mathematically
well defined theory of collectives. However, the present paper is
far away all those problems with the notion of randomness. All our
considerations are related only to the statistical stabilization
of relative frequencies. We do not take care on randomness. We
belief that all sequence induced by e.g. quantum statistical
experiments are random.

Mathematical formalization of the notion of context in general
case is a problem of large complexity. In this paper we propose
the following definition of {\it quantum conmtext.}

{\bf Definition.} Every family  ${\cal A}=\{A_1, A_2,...\}$
(finite or infinite) of self-adjoint commutative operators is said
to be a quantum context.

{\bf Example 1.} ( Space-time context). Let  ${\cal A}=\{A_1,
A_2,A_3,A_4\}$ be the system of generators of the unitary group of
translations. Then ${\cal A}$ is said to be the space-time
context.

{\bf Example 2.} (Internal symmetry). Let $G$ by a compact Lie
group if internal symmetries (for example, the gauge group $U(1)$
which describes the electric charge ). Then generators of the
unitary representation of the group defines the internal symmetry
context.

\section{Bell`s Theorem and Stochastic Processes}

In the  presentation of Bell's theorem we will follow ~\cite{Vol1}
where one can find also more references. {\it Bell's theorem,} as
it is formulated  in \cite{Vol1},  reads:
\begin{equation}
\label{BT}
(a, b) \neq E\xi (a) \eta(b) \label{eq:eqn1}
\end{equation}
were $a=(a_1,a_2,a_3)$ and $b=(b_1,b_2,b_3)$ are two unit vectors
in three-dimensional space ${\bf R}^3$. Here $\xi(a)= \xi
(a,\lambda)$ and $ \eta (b)= \eta(b,\lambda)$ are random  fields
on the sphere, $\lambda$ is an element from the probability space
 $(\Lambda,
{\cal F}, d\rho (\lambda))$. Here $\Lambda$ is a set, ${\cal F}$
is a sigma-algebra of subsets and $d\rho (\lambda)$ is a
probability measure, i.e. $d\rho (\lambda) \geq 0,~\int d\rho
(\lambda)=1.$ The expectation is
$$
Ef=\int_{\Lambda}f(\lambda)d\rho (\lambda)
$$
The random fields satisfy the bound
\begin{equation}
|\xi (a,\lambda)|\leq 1,~  | \eta (b,\lambda)|\leq 1
 \label{Bou}
\end{equation}
The theorem says that there exists no probability space  and a
pair of stochastic processes with indicated properties such that
their expectation is equal to the scalar product of the vectors
$a$ and $b$. The form (\ref{BT}) is convenient for various
generalizations. o

 Let us discuss now a physical
interpretation of this result. In the Bohm formulation of the EPR
argument one considers a pair of spin one-half particles formed in
the singlet spin state and moving freely towards two detectors. If
one neglects the space part of the wave function  then one has the
Hilbert space $C^2\otimes C^2$ and  the quantum mechanical
correlation of two spins in the singlet state $\psi_{spin}\in
C^2\otimes C^2$ is
\begin{equation}
 D_{spin}(a,b)=\left<\psi_{spin}|\sigma\cdot a \otimes\sigma\cdot
b|\psi_{spin}\right>=-a\cdot b \label{eq:eqn1}
\end{equation}
Here $a=(a_1,a_2,a_3)$ and $b=(b_1,b_2,b_3)$ are two unit vectors
in three-dimensional space ${\bf R}^3$,
$\sigma=(\sigma_1,\sigma_2,\sigma_3)$ are the Pauli matrices, $
\sigma\cdot a =\sum_{i=1}^{3}\sigma_i a_i$
 and
$$
\psi_{spin}=\frac{1}{\sqrt 2} \left(\left(
    \begin{array}{c}0\\1
    \end{array}
    \right)
\otimes \left(
    \begin{array}{c}1\\
    0\end{array}
    \right)
-\left(
    \begin{array}{c}1\\
    0\end{array}
    \right)
\otimes \left(
    \begin{array}{c}0\\
    1\end{array}
    \right)
\right).
$$
 The proof of
the theorem is based on Bell`s or the Clauser-Horn-Shimony-Holt
(CHSH) inequalities. Let us stress that the main point in the
mathematical proof is actually not the discretness of the
classical or quantum spin variables and even not a nonlocality but
the bound (\ref{Bou}) for classical random fields.

\subsection{Classical model of spin correlation}

To explain the last point we present here a simple local in the
sense of Bell classical probabilistic model which reproduces the
quantum mechanical correlation of two spins. Let us take as a
probability space $\Lambda$ just 3 points: $\Lambda =\{1,2,3\}$
and the expectation
$$
Ef=\frac{1}{3}\sum_{\lambda =1}^3f(\lambda)
$$
Let the random fields be
$$
\xi (a,\lambda)=\eta (a,\lambda)=\sqrt 3 a_\lambda,~~
\lambda=1,2,3
$$
Then one has the relation:
$$
(a,b)=E\xi(a)\xi(b)
$$
The Bell`s theorem (\ref{BT}) does not valid in this case because
we do not have the bound (\ref{Bou}). Instead we have
$$
|\xi (a,\lambda)|\leq \sqrt 3
$$
This model shows that the bound (\ref{Bou}) plays the crucial role
in the proof of Bell`s theorem.  Actually to reproduce $(a,b)$ we
can use even a deterministic model: simply the first
experimentalist will report about the measurement of the
components of the vector $(a_1,a_2,a_3)$ and the second about the
measurement of the components of the vector $(b_1,b_2,b_3)$.
\subsection{CHSH Inequality}

The proof of Bell's theorem is based on the following theorem
which is a slightly generalized the Clauser-Horn-Shimony-Holt
(CHSH) result.

{\bf Theorem 2.} Let $f_1,~f_2,~g_1$ and $g_2$ be random variables
(i.e. measured functions) on the probability space $(\Lambda,
{\cal F}, d\rho (\lambda))$ such that
\begin{equation}
\label{eq:Ab2} |f_i(\lambda)g_j(\lambda)|\leq 1,~~i,j=1,2.
\end{equation}
Denote
$$
P_{ij}=Ef_ig_j,~~i,j=1,2.
$$
Then
\begin{equation}
\label{eq:Ab3}|P_{11}-P_{12}|+|P_{21}+P_{22}|\leq 2
\end{equation}

The last inequality is called
  the CHSH inequality.

\section{Correlation functions in EPR model}

Now let us apply  similar approach to the original EPR case
\cite{KhV2}. The Hilbert space of two one-dimensional particles is
$L^2(R)\otimes L^2(R)$ and canonical coordinates and momenta are
$q_1,q_2,p_1,p_2$ which obey the commutation relations
\begin{equation}
\label{eq:EPR1}
[q_m,p_n]=i\delta_{mn},~~[q_m,q_n]=0,~~[p_m,p_n]=0,~~m,n=1,2
\end{equation}

The EPR paradox can be described as follows. There is such a state
of two particles that  by measuring $p_1$ or $q_1$ of the first
particle, we can predict with certainty and without interacting
with the second particle, either the value of $p_2$ or the value
of $q_2$ of the second particle. In the first case $p_2$ is an
element of physical reality, in the second $q_2$ is. Then, these
realities must exist in the second particle before any measurement
on the first particle since it is assumed that the particle are
separated by a space-like interval. However the realities can not
be described by quantum mechanics because they are incompatible --
coordinate and momenta do not commute. So that EPR conclude that
quantum mechanics is not complete. Note that the EPR state
actually is not a normalized state since it is represented by the
delta-function, $\psi=\delta(x_1-x_2-a).$

An important point in the EPR consideration is that one can choose
what we measure -- either the value of $p_1$ or the value of
$q_1$.

For a mathematical formulation of a free choice we introduce
canonical transformations of our variables:
\begin{equation}
\label{eq:EPR2} q_n(\alpha)=q_n\cos \alpha - p_n\sin \alpha,~~
p_n(\alpha)=q_n\sin\alpha + p_n\cos\alpha;~~n=1,2
\end{equation}
Then one gets
\begin{equation}
\label{eq:EPR3} [q_m(\alpha),p_n(\alpha)]=i\delta_{mn};~~n=1,2
\end{equation}
In particular one has $q_n(0)=q_n,~~q_n(3\pi/2)=p_n,~~n=1,2.$

Now let us consider the correlation function
\begin{equation}
\label{eq:EPR4} D(\alpha_1,\alpha_2)=\left<\psi|q_1 (\alpha_1)
\otimes q_2(\alpha_2)|\psi\right>
 \end{equation}
The correlation function $D(\alpha_1,\alpha_2)$ (\ref{eq:EPR4}) is
an analogue of the Bell correlation function  $D_{spin}(a,b)$
(\ref{eq:eqn1}). Bell in \cite{Bel2} has suggested to consider the
correlation function of just the free evolutions of the particles
at different times (see  below).

 We are interested in the
question whether the quantum mechanical correlation function
(\ref{eq:EPR4}) can be represented in the form
\begin{equation}
\label{eq:EPR5} \left<\psi|q_1 (\alpha_1) \otimes
q_2(\alpha_2)|\psi\right>=E \xi_1 (\alpha_1) \xi_2(\alpha_2)
\end{equation}
Here $ \xi_n (\alpha_n)=\xi_n (\alpha_n,\lambda), n=1,2$ are two
real  random processes, possibly unbounded The parameters
$\lambda$ are interpreted as hidden variables in a realist theory.

{\bf Theorem}.
 For an arbitrary state $\psi\in
L^2(R)\otimes L^2(R)$ on which products of operators
$q_1,q_2,p_1,p_2$ are defined there exist random processes $\xi_n
(\alpha_n,\lambda)$ such that the relation (\ref{eq:EPR5}) is
valid.

{\bf Proof.} We rewrite the correlation function
$D(\alpha_1,\alpha_2)$ (\ref{eq:EPR4}) in the form
\begin{equation}
\label{eq:EPR6} \left<\psi|q_1 (\alpha_1) \otimes
q_2(\alpha_2)|\psi\right>=
<q_1q_2>\cos\alpha_1\cos\alpha_2-<p_1q_2>\sin\alpha_1\cos\alpha_2
\end{equation}
$$
-<q_1p_2>\cos\alpha_1\sin\alpha_2
+<p_1p_2>\sin\alpha_1\sin\alpha_2
$$
Here we use the notations as
$$<q_1q_2>=\left<\psi|q_1
q_2|\psi\right>$$
Now let us set
$$
\xi_1(\alpha_1,\lambda)=f_1(\lambda)\cos\alpha_1-g_1(\lambda)\sin\alpha_1,
$$
$$
\xi_2(\alpha_2,\lambda)=f_2(\lambda)\cos\alpha_2-g_2(\lambda)\sin\alpha_2
$$
Here real functions $f_n(\lambda),g_n(\lambda),~n=1,2$ are  such that
\begin{equation}
\label{eq:EPR7}
Ef_1f_2=<q_1q_2>,~Eg_1f_2=<p_1q_2>,~Ef_1g_2=<q_1p_2>,~Eg_1g_2=
<p_1p_2>
\end{equation}
We use for the expectation the notations as
$Ef_1f_2=\int f_1(\lambda) f_2,(\lambda) d\rho(\lambda)$. To solve
the system of equations (\ref{eq:EPR7}) we take
\begin{equation}
\label{EPR7a}
f_n(\lambda)=\sum_{\mu
=1}^2F_{n\mu}\eta_{\mu}(\lambda),~ g_n(\lambda)=\sum_{\mu
=1}^2G_{n\mu}\eta_{\mu}(\lambda)
\end{equation}
where $F_{n\mu},G_{n\mu}$ are constants and
$E\eta_{\mu}\eta_{\nu}=\delta_{\mu\nu}$. We denote
$$
<q_1q_2>=A,~<p_1q_2>=B,~<q_1p_2>=C,~<p_1p_2>=D.
$$
A solution of Eqs (\ref{eq:EPR7}) may be given for example by
$$
f_1=A\eta_1,~~f_2=\eta_1,
$$
$$
g_1=B\eta_1+(D-\frac{BC}{A})\eta_2,~g_2=\frac{C}{A}\eta_1+\eta_2
$$
Hence the representation of the quantum correlation function in
terms of the separated classical random processes  (\ref{eq:EPR5})
is proved.

{\bf Remark 1.} We were able to solve the system of equations
(\ref{eq:EPR7}) because there are no bounds to the random
variables $f_1,f_2,g_1,g_2.$ In the case of the Bohm spin model
one has the bound (\ref{eq:Ab2}) which leads to the CSHS
inequality (\ref{eq:Ab3}) and as a result an analogue of equations
(\ref{eq:EPR7}) in the Bohm model has no solution.

{\bf Remark 2.} The condition of reality of the functions $ \xi_n
(\alpha_n,\lambda)$ is important. It means that the range of $
\xi_n (\alpha_n,\lambda) $ is the set of eigenvalues of the
operator $q_n(\alpha_n).$ If we relax this condition then one can
get a hidden variable representation just by using an expansion of
unity:
$$
\left<\psi|q_1 (\alpha_1)
q_2(\alpha_2)|\psi\right>=\sum_{\lambda}\left<\psi|q_1 (\alpha_1)|\lambda\right>
\left<\lambda|q_2(\alpha_2)|\psi\right>
$$
For a discussion of this point in the context of a noncommutative
spectral theory see \cite{Vol2}.

Similarly one can prove a representation
\begin{equation}
\label{eq:EPR8} \left<\psi|q_1 (t_1) \otimes
q_2(t_2)|\psi\right>=\int \xi_1 (t_1,\lambda) \xi_2(t_2,\lambda)
d\rho(\lambda)
\end{equation}
where $q_n(t)=q_n+p_nt,~n=1,2$ is a free quantum evolution of the
particles. It is enough to take
$$
\xi_1(t_1,\lambda)=f_1(\lambda)+g_1(\lambda)t_1,
~~\xi_2(t_2,\lambda)=f_2(\lambda)+g_2(\lambda)t_2.
$$
{\bf Remark 3.} In fact we can prove a more general theorem. If
$f(s,t)$ is a function of two variables then it can be represented
as the expectation of two stochastic processes:
$f(s,t)=E\xi(s)\eta(t). $ Indeed, if $ f(s,t)=\sum_n g_n(s)h_n(t)
$ then we can take
$$
\xi(s,\omega)=\sum_ng_n(s)x_n(\omega),~\eta(t,\omega)=
\sum_nh_n(s)x_n(\omega)
$$
where $ Ex_nx_m=\delta_{nm}. $

\section {Space-time dependence of correlation
functions and disentanglement}

\subsection{Modified Bell`s equation}

In the previous sections the space part of the wave function of
the particles was neglected. However exactly the space part is
relevant to the discussion of locality. The Hilbert space assigned
to one particle with spin 1/2 is  ${\bf C}^2\otimes L^2({\bf
R}^3)$ and the Hilbert space of two particles is ${\bf C}^2\otimes
L^2({\bf R}^3)\otimes {\bf C}^2\otimes L^2({\bf R}^3).$ The
complete wave function is $\psi =(\psi_{ij}({\bf r}_1,{\bf
r}_2,t))$ where $i$ and $j $ are spinor indices, $t$ is time  and
${\bf r}_1$ and ${\bf r}_2$ are vectors in three-dimensional
space.

We suppose that there are two  detectors (A and B) which are
located in space ${\bf R}^3$ within the two localized regions
${\cal O}_1$ and ${\cal O}_2$ respectively, well separated from
one another. If one makes a local observation in the region ${\cal
O}_1$ then this means that one measures not only the spin
observable $\sigma_i$ but also some another observable which
describes the localization of the particle like the energy density
or the projection operator $P_{{\cal O}}$ to the region ${\cal
O}$. Normally in experiments there are polarizers  and detectors.
We will consider here correlation functions which includes the
projection operators $P_{{\cal O}}$.

Quantum correlation describing the localized measurements of spins
in the regions ${\cal O}_1$ and ${\cal O}_2$ is

\begin{equation}
\label{eq:eqn6} \omega(\sigma\cdot a   P_{{\cal O}_1}\otimes
\sigma\cdot b  P_{{\cal O}_2})=\left<\psi| \sigma\cdot a P_{{\cal
O}_1}\otimes  \sigma\cdot b  P_{{\cal O}_2} |\psi\right>
\end{equation}

Let us consider the simplest case when the wave function has the
form of the product of the spin function and the spacial function
$\psi=\psi_{spin}\phi({\bf r}_1,{\bf r}_2)$. Here $\phi({\bf
r}_1,{\bf r}_2)$ is a complex valued function. Then one has
\begin{equation}
\label{eq:eqn7}
 \omega(\sigma\cdot a   P_{{\cal O}_1}\otimes
\sigma\cdot b  P_{{\cal O}_2})=
 =g ({\cal O}_1,{\cal O}_2)
  D_{spin}(a,b)
\end{equation}
where the function
\begin{equation}
\label{eq:eqn8}
 g ({\cal O}_1,{\cal O}_2)=\int_{{\cal O}_1 \times {\cal O}_2}|\phi({\bf
 r}_1,{\bf
 r}_2)|^2 d{\bf r}_1d{\bf r}_2
\end{equation}
describes correlation of particles in space. It is the probability
to find one particle in the region ${\cal O}_1$ and another
particle in the region ${\cal O}_2$.

One has
\begin{equation}
\label{eq:eqn8g} 0\leq g ({\cal O}_1,{\cal O}_2)\leq 1
\end{equation}

\subsection{ Disentanglement}

If ${\cal O}_1$ is a bounded region and ${\cal O}_1(l)$ is a
translation of ${\cal O}_1$ to the 3-vector $l$ then one can prove

\begin{equation}
\label{eq:eqn8l} \lim_{|l|\to\infty} g({\cal O}_1(l),{\cal O}_2)=0
\end{equation}

Since
$$\left<\psi_{spin}|\sigma\cdot a \otimes
I|\psi_{spin}\right>=0
$$
we have
$$
\omega (\sigma\cdot a P_{{\cal O}_1}\otimes I)=0.
$$
Therefore we have proved the following proposition which says that
the state  $\psi=\psi_{spin}\phi({\bf r}_1,{\bf r}_2)$ becomes
disentangled (factorized) at large distances.

{\bf Proposition.} One has the following property of the
asymptotic factorization (disentanglement) at large distances:
 \begin{equation}
\label{eq:eqn8ld} \lim_{|l|\to\infty} [\omega (\sigma\cdot a
P_{{\cal O}_1(l)}\otimes \sigma\cdot b P_{{\cal O}_2})- \omega
(\sigma\cdot a P_{{\cal O}_1(l)}\otimes I )\omega(I\otimes
\sigma\cdot b P_{{\cal O}_2} )]=0
\end{equation}
or
$$
\lim_{|l|\to\infty} \omega (\sigma\cdot a P_{{\cal O}_1(l)}\otimes
\sigma\cdot b P_{{\cal O}_2})=0.
$$
Now one inquires whether one can write a representation
\begin{equation}
\label{eq:eqn9}
 \omega(\sigma\cdot a   P_{{\cal O}_1}\otimes
\sigma\cdot b  P_{{\cal O}_2})=
 \int \xi_1 (a,{\cal O}_1,\lambda)
 \xi_2 (b,{\cal O}_2,\lambda) d\rho(\lambda)
\end{equation}
where $|\xi_1 (a,{\cal O}_1,\lambda)|\leq 1,~~ |\xi_2 (b,{\cal
O}_2,\lambda)|\leq 1$.

{\bf Remark.} A local modified equation reads
$$
|\phi ({\bf r_1},{\bf r_2},t)|^2(a,b) =E\xi (a,{\bf r_1},t) \eta
(b,{\bf r_2},t).
$$

If we are interested in the conditional probability of finding the
projection of spin along  vector $a$ for the particle 1  in the
region ${\cal O}_1$ and the projection of spin along the vector
$b$ for the particle 2 in the region ${\cal O}_2$   then we have
to divide both sides of Eq.~(\ref{eq:eqn9}) by $g({\cal O}_1,{\cal
O}_2)$.

Note  that here the classical random variable $\xi_1=\xi_1
(a,{\cal O}_1,\lambda)$ is not only separated in the sense of Bell
(i.e. it depends only on $a$) but it is also local in the 3 dim
space since it depends only on the region ${\cal O}_1$. The
classical random variable $\xi_2$ is also local in 3 dim space
since it depends only on ${\cal O}_2$. Note also that since the
eigenvalues of the projector $P_{{\cal O}}$ are 0 or 1 then one
should have $ |\xi_n (a,{\cal O}_n)|\leq 1,~n=1,2.$

Due to the property of the asymptotic factorization and the
vanishing of the quantum correlation for large $|l|$ there exists
a trivial asymptotic classical representation of the form
(\ref{eq:eqn9}) with $\xi=\eta=0.$

We can do even better and find a classical representation which
will be valid uniformly for large $|l|$.

Let us take now the wave function $\phi$ of the form
$\phi=\psi_{1}({\bf r}_1)\psi_{2}({\bf r}_2)$  where
$$
\int_{R^3}|\psi_{1}({\bf r}_1)|^2d{\bf r}_1=1,~~
\int_{R^3}|\psi_{2}({\bf r}_2)|^2d{\bf r}_2=1
$$
In this case
$$
 g ({\cal O}_1(l),{\cal O}_2)=
 \int_{{\cal O}_1(l)}|\psi_{1}({\bf r}_1)|^2d{\bf r}_1\cdot
 \int_{{\cal O}_2}|\psi_{2}({\bf r}_2)|^2d{\bf r}_2
$$
There exists such $L>0$ that
$$
\int_{B_L}|\psi_{1}({\bf r}_1)|^2d{\bf r}_1=\epsilon <1/2,~~
$$
where $B_L=\{{\bf r}\in R^3: |{\bf r}|\geq L\}.$

 We have the
following

{\bf Theorem 4.} Under the above assumptions and for large enough
$|l|$ there exists the following representation of the quantum
correlation function
$$
\omega (\sigma\cdot a P_{{\cal O}_1(l)}\otimes \sigma\cdot b
P_{{\cal O}_2})
 =E\xi ({\cal O}_{1}(l),a)  \xi({\cal O}_2,b)
 $$
 where all classical random variables are bounded by 1.

\section{Conclusions}

Mathematical definitions of local realism in the sense of Bell and
in the sense of Einstein are given in the paper. We show how the
space-time structure can be considered from the contextual point
of view. A mathematical framework for the contextual approach is
outlined. We demonstrate that if we include into the quantum
mechanical formalism the space-time structure in the standard way
then quantum mechanics might be consistent with Einstein's local
realism. It shows that loopholes are unavoidable in experiments
aimed to establish a violation of Bell`s inequalities.

It is shown also that, in contrast to the Bell`s theorem for the
spin or polarization variables, for the original EPR correlation
functions which deal with positions and momenta one can get a
local realistic representation in terms of separated random
processes. The representation is obtained for any state including
entangled states. Therefore the original EPR model does not lead
to quantum nonlocality in the sense of Bell even for entangled
states. One can get quantum nonlocality in the EPR situation only
if we rather artificially restrict ourself in the measurements
with a two dimensional subspace of the infinite dimensional
Hilbert space corresponding to the position or momentum
observables.
 An interrelation of the roles of entangled states
and the bounded observables in  considerations of local realism
and quantum nonlocality deserves a further study.

 It is important
 to develop further the mathematical theory of context in classical
and in quantum theory.

 { \bf
Acknowledgments}

The paper was supported by the grant of The Swedish Royal Academy
of Sciences on collaboration with scientists of former Soviet
Union, EU-network on quantum probability and applications and
RFFI-0201-01084, INTAS-990590.

\end{document}